# Fiber-based Ultra-High Speed Diffuse Speckle Contrast Analysis System for Deep Blood Flow Sensing Using a Large SPAD Camera

Quan Wang[1], Renzhe Bi[2], Songhua Zheng[2], Ahmet T. Erdogan[3], Yi Qi[2], Chenxu Li[1], Yuanyuan Hua[3], Mingliang Pan[1], Yining Wang[3], Neil Finlayson[3], Malini Olivo[2], Robert K. Henderson[3], and David Uei-Day Li[1, a]

**AFFILIATIONS**

[1]University of Strathclyde, Faculty of Engineering, Department of Biomedical Engineering, Glasgow, UK.

[2]The A* STAR Skin Research Labs (A* SRL), Agency for Science, Technology and Research (A* STAR), 31 Biopolis Way, Nanos, Singapore 138669, Republic of Singapore

[3]The University of Edinburgh, School of Engineering, Integrated Nano and Micro Systems (IMNS), Edinburgh, EH9 3JL, UK.

[a] **Author to whom correspondence should be addressed:** david.li@strath.ac.uk

**ABSTRACT**

Diffuse speckle contrast analysis (DSCA), also called speckle contrast optical spectroscopy (SCOS), has emerged as a groundbreaking optical imaging technique for tracking dynamic biological processes, including blood flow and tissue perfusion. Recent advancements in single-photon avalanche diode (SPAD) cameras have unlocked exceptional capabilities in sensitivity, time resolution, and high frame-rate imaging. Despite this, the application of large-format SPAD arrays in speckle contrast analysis is still relatively uncommon. In this study, we introduce a pioneering use of a large-format SPAD camera for DSCA. By harnessing the camera's high temporal resolution and photon-detection efficiency, we significantly enhance the accuracy and robustness of speckle contrast measurements. Our experimental results demonstrate the system's remarkable ability to capture rapid temporal variations over a broad field of view, enabling detailed spatiotemporal analysis. Through simulations, phantom experiments, and in vivo studies, we validate the approach's potential for a wide range of biomedical applications, such as cuff occlusion tests and functional tissue monitoring. This work highlights the transformative impact of large SPAD cameras on DSCA, setting the stage for new breakthroughs in optical imaging.

# I. INTRODUCTION

In a healthy individual, proper blood flow (BF) is vital for maintaining a steady supply oxygen and essential energy sources, such as glucose and lactate, to organs, while also ensuring the efficient removal of metabolic waste products [1]. Of particular importance is cerebral blood flow (CBF), which is crucial for optimal brain function [2], brain metabolism [3], and the brain's ability to respond to external stimuli [4]. For adults, typical CBF is around 50 ml/(100 g min) [5], whereas in newborns, it ranges between 10-30 ml/(100 g min) [6]. Any disruption to CBF can result in significant brain damage, including ischemic injury or stroke [7]. Real-time monitoring of blood flow is essential for diagnosing and managing a variety of medical conditions, such as stroke, traumatic or hypoxic-ischemic encephalopathy (HIE) [8], neurological disorders, cardio-cerebral diseases, cancer treatments, tissue perfusion in peripheral vascular diseases [9], brain health and function [10], wound healing, sepsis, shock [11], skeletal muscle injuries [12], and tissue viability during surgical procedures.

Several optical methods have been employed for non-invasive monitoring of both healthy and pathophysiological tissues, including laser speckle contrast imaging (LSCI) [13], laser Doppler flowmetry (LDF), diffuse correlation spectroscopy (DCS), and Diffuse speckle contrast analysis (DSCA) – also referred to as speckle visibility spectroscopy (SVS) or speckle contrast optical spectroscopy (SCOS). A key distinction between these techniques lies in their penetration depth: while LSCI and LDF are typically limited to superficial tissues (less than 1 mm) due to their reliance on single or few dynamic scattering events, DSCA/SCOS and DCS can penetrate much deeper, reaching several centimeters into tissue.

DCS relies on temporal sampling methods, where traditional avalanche photodiodes (APDs) or advanced single-photon avalanche diode (SPAD) detectors capture intensity fluctuations from one or a few speckle grains to reconstruct temporal dynamics. In contrast, DSCA uses a spatial sampling approach, which doesn't require a detector with a high frame rate. Instead, the camera typically operates with an exposure time longer than the speckle field's decorrelation time, utilizing a larger detection area with many pixels to capture more photons and speckles. Originally, Bi *et al.* first proposed DSCA [14–16], drew heavily from the concepts of LSCI, focusing on average values rather than imaging blood flow. DSCA is sometimes referred to as SVS [17] or SCOS [18], and it has been extensively studied theoretically [19,20] and experimentally [21,22]. Notably, Kim *et al.* (2023) demonstrated that DSCA/SCOS outperforms DCS, offering more than a 10-fold improvement in signal-to-noise ratio (SNR) at a comparable cost. Importantly, while the setup for a fiber-based DSCA system is identical to that of DCS, DSCA does not require model-based fitting to separate absorption and scattering effects from the dynamic signal, and thus it does not provide a quantitative estimate of blood flow [18]. Conventionally, DSCA combines the deep tissue penetration capabilities of DCS with the relatively low-cost CCD or CMOS detectors used in LSCI. However, despite the use of these affordable detectors, DSCA has not yet seen widespread commercial adoption. With the introduction of faster photon-counting detectors, such as SPAD cameras, which have negligible readout noise, it is now possible to apply the spatial DSCA technique with improved statistical reliability in speckle contrast calculations.

In this study, we introduce a novel, compact 512 × 512 pixel SPAD array called ATLAS [23,24] built with industry-standard CMOS technology, for deep tissue blood flow (BF) measurements using the DSCA approach. This detector array boasts high photon efficiency, minimal dead time, zero readout noise, and a high frame rate (up to 27kfps), all of which are essential for precise BF measurements in low-light, in vivo conditions. Future advancements could enable real-time BF monitoring in freely moving subjects. To evaluate our ATLAS-DSCA system, we conducted deep tissue BF measurements in four healthy volunteers during arterial arm cuff occlusion and forehead BF monitoring. Additionally, vibration phantom experiments were performed, and the results were compared to those from a traditional DSCA system.

A summary of existing DSCA/SCOS systems is provided in Table I, detailing laser wavelength, applications, sampling rate, source-detector separation, and sensor type. This table highlights the evolution of DSCA technology and contextualizes the significance of our proposed system in advancing the field.

The structure of this paper is as follows: we first present the theoretical principles of DSCA, followed by a description of the SPAD camera architecture and system implementation. Next, we compare our phantom results with those from traditional DSCA methods. Finally, we present the in vivo experimental results, discuss the advantages and limitations of the SPAD-DSCA system, and propose potential future improvements.

Table I Existing DSCA/SCOS system

| Laser | Wavelength (nm) | Applications | Sampling rate (Hz) | Fiber-based/fiberless | Source-detector separation (mm) | System name | Year | Sensor | Ref. |
|---|---|---|---|---|---|---|---|---|---|
| CW | 785 | Forearm | 30 | Fiber-based | 24 | DSCA | 2013 | EMCCD | 14 |
| CW | 785 | Forearm and palm | 1 | Fiber-based | 15 | tDSCA | 2013 | CCD | 15 |
| CW | 785 | Forearm | N.A. | Fiberless | 30 | SCOS | 2014 | CCD | 18 |
| CW | 671 | Phantom | N.A. | Fiberless | 18 | DSCA | 2017 | CCD | 25 |
| CW | 785 | Forearm & forehead | N.A. | Fibreless | 20 | SCOS | 2018 | SPAD (32×2) | 26 |
| CW | 785 | Forearm, & forehead | 300 | Fiber-based | 25 | DSCA | 2020 | CCD | 16 |
| CW | 785 | Forehead | N.A. | Fiber-based | 26 | SCOS | 2023 | sCMOS | 27 |
| VHG holographic | 852 | Forearm, forehead & arithmetic tests | N.A. | Fiber-based | 45 | SCOS | 2023 | sCMOS | 28 |
| CW | 785 | Forehead | 80 | Compact and fiberless | 50 | SCOS | 2024 | Sony IMX392 | 29 |
| CW | 785 | Diabetic | 330 | Fiber-based | 12 | DSCA | 2024 | CCD | 30 |
| CW | 785 | Forearm | N.A. | Fiber-based | 25 | DSCA | 2024 | Generic photodiode | 31 |
| CW | 808 | Wrist, exercise | 390 | Fiber-based | 4.5 | SCOS | 2024 | Basler boost | 32 |
| CW | 785 | Forearm, forehead & arithmetic tests | > 800 | Fiber-based | 30 | DSCA | 2025 | SPAD (512×512) | ours |

## II. METHODS

### A. Theoretical Background

DSCA relies on the speckle contrast ($\kappa$), defined as the ratio of the standard deviation ($\sigma_I(\rho, T)$) of the measured intensity during a specific exposure time to its mean ($\langle I \rangle$) across various speckles. DSCA quantifies the statistics of the fluctuation of the speckle pattern as the variance of the measured intensity ($\sigma_I^2$) either in spatial or temporal domains [33]:

$$\kappa^2(\rho, T) = \frac{\sigma_I^2(\rho,T)}{\langle I(\rho,T)\rangle^2} \quad (1)$$

here $\rho$ denotes the source-detector separation, $T$ is the exposure time, and $\kappa^2$ varies between zero and one, with a higher value indicating a slower scatterer fluctuation. $\kappa$ is related to the normalized electric field auto-correlation function ($g_1(r,T)$) is given by:

$$\kappa^2(\rho, T) = \frac{2\beta}{T}\int_0^T (1 - \frac{\tau}{T})|g_1(\rho,\tau)|^2 d\tau. \quad (2)$$

where $g_1(\rho, \tau) = G_1(\rho,\tau)/G_1(\rho, 0)$ and $G_1(\rho, \tau)$ represents the Green's function of Brownian motion at $\rho$ in a semi-infinite geometry [34]. In this equation, $\beta$ is an experimental constant that accounts for the collection optics.

### B. Noise Correction

In practical applications, it is crucial to adjust the speckle contrast calculation to account for shot noise and other noise sources inherent in the detection system. Since speckle contrast ($\kappa$) is influenced by the variance from the expected theoretical behavior, deviations become particularly pronounced in regions with lower signal-to-noise ratio (SNR). It is important to note that the calculation in Eq. (1) does not account for these additional noise contributions, especially in areas with lower SNR. To address this, we define a corrected squared speckle contrast $\kappa_c^2$ as follows:

$$\kappa_c^2 = \kappa_{measured}^2 - \kappa_{shot}^2 - \kappa_{dark}^2 \quad (3)$$

Before calculating $\kappa$, the raw intensity images are corrected by subtracting dark counts and removing bad pixels. The dark offset, determined as the average of 1000 dark images (with the laser off), is subtracted from the raw intensity to obtain the corrected intensity $I_c = I - I_D$. Even though the intensity is corrected, the variance from the dark noise is included in the intensity's variance, making it essential to deduct the dark variance $\sigma_D^2$. Another significant noise source is inherent shot noise, which follows Poisson statistics, with a variance equal to the mean intensity in electrons [e-], defined as $\sigma_s^2 = \langle I_c \rangle$. Since SPAD cameras have no readout noise, Eq. (3) becomes:

$$\kappa_c^2 = \frac{\sigma_{I_c}^2 - \sigma_D^2 - \sigma_s^2}{\langle I_c \rangle^2}. \quad (4)$$

The blood flow index (BFi) is then related to $\kappa_c^2$ by:

$$BFi = \frac{1}{\kappa_c^2} \quad (5)$$

In all the findings presented in this paper, we use the normalized blood flow index (normalized BFi) to provide standardized BF data, enhancing comparability across measurements. The BFi metric reflects the total blood volume transported within a specific time frame. According to Poiseuille's law [35], BF is strongly influenced by factors such as blood pressure, vessel radius, viscosity, and vessel length. Even small changes in vessel radius can have a significant impact on BF due to its fourth-power relationship with radius.

### C. SPAD (ATLAS) Architecture

The sensor, called ATLAS hereafter, features a 512 × 512 array of deep trench isolation (DTI) microlensed SPADs with a 10.17 µm pitch. It offers a peak photon detection efficiency (PDE) of 55% (26% at 940 nm) and a median dark count rate (DCR) of 500 cps at room temperature, operating at 23 V with a breakdown voltage of 17.8 V [36]. The passively quenched SPADs are organized into 4×4 groups using an OR-tree structure to form 128 × 128 macropixels, each with a 40.68 µm pitch. Each macropixel is capable of computing a 31-tap autocorrelation function with a minimum correlation time of 1 µs (which can be reduced to 100 ns, depending on the clock rate).

ATLAS supports the following operating modes:

1. A 22-bit single-photon counting mode.

2. A time-gated 22-bit single-photon counting mode via a balanced H-tree, with time gates generated by an FPGA or on-chip DLL at 15–30 ps granularity.

3. A multispeckle DCS mode, delivering the average of individual pixel autocorrelations in an on-chip, high-speed "ensemble" mode [23,24].

4. A DCS imaging mode [23,24].

In ATLAS, each photon is directly converted into a 1-bit count, eliminating the readout noise typically encountered in CCD or CMOS imagers. The clock frequency is adjustable (e.g., 20 MHz, 25 MHz, 50 MHz, 75 MHz), and TBIN_CLK_PERIODS can range from 32 to 65,535. The chip is mounted on a PCB board and connected to a field-programmable gate array (FPGA), such as the Opal Kelly 7310-A200.

In the photon counting (PC) mode, as shown in FIG. 1(a), each pixel acquires photons during a single exposure (TBIN period) and photons are counted with a 5-bit ripple counter. At the end of the TBIN exposure, the shift register is clocked, transferring the 5-bit count into the first element (C(τ₀)) and subsequently accumulated to the first SRAM address. This is repeated for a user defined number of integration cycles (i.e., TINT_TBIN_ITERATIONS). After which the accumulated 22-bit count is read out via the column bus. It is worth noting that in the PC mode only the first elements of the Shift Register and SRAM are utilised, and the Multiplier is configured to multiply C(τ₀) with 1 (effectively by-passing the Multiplier). The remaining 31 shift register and SRAM elements are utilised in other DCS related modes. FIG. 1(b) shows the Global shutter timing diagram for photon counting mode.

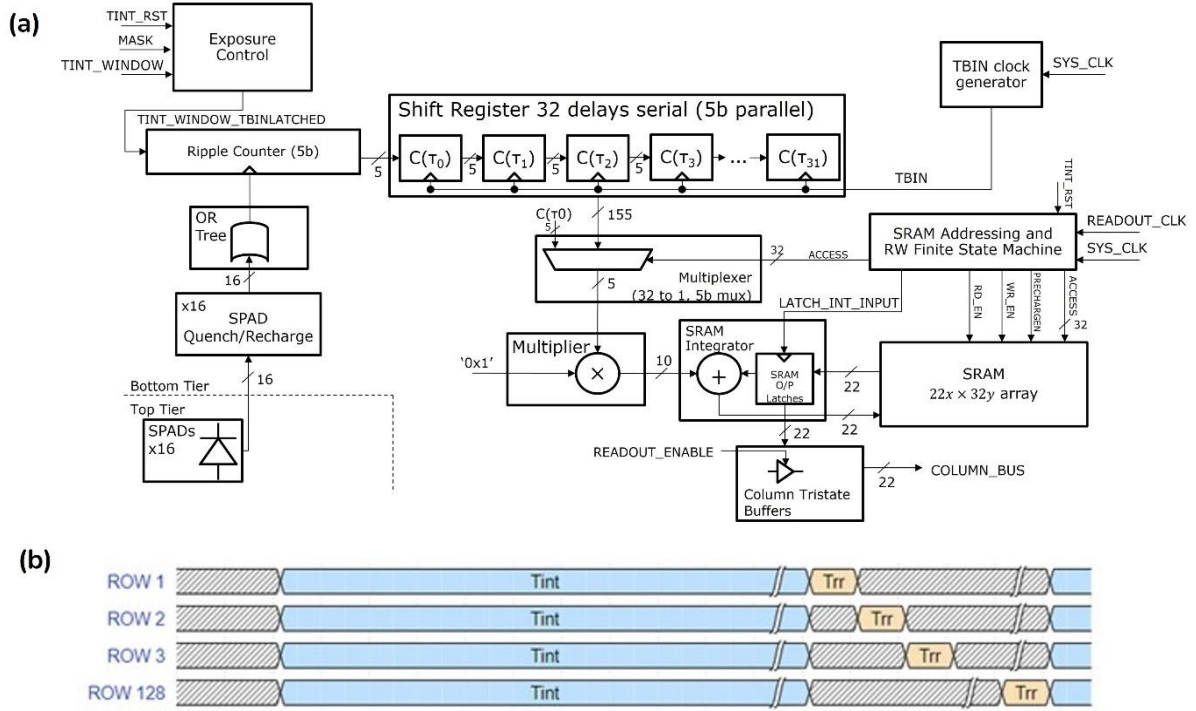

FIG. 1. (a) Block diagram showing main blocks of the pixel schematic operating in Photon Counting mode. (b) Global shutter timing diagram for photon counting mode. Integration time: Tint, Row readout time: Trr.

**D. Implementation（Experimental Setup）**

The experimental setup is schematically illustrated in FIG. 2(a). A long-coherence 785-nm laser (>5 m coherence length, DL785-100-S from CrystaLaser) serves as the light source. Laser light is delivered to the phantom through a multimode (MM) fiber with a 200 μm core, while scattered light is collected by another MM fiber with the same core size. One end of the detection fiber is placed on the phantom's surface, with the other end aligned to the center of the SPAD camera. Zoomed-in images of the SPAD chip (front view) and the Opal Kelly FPGA board (back view) are also provided.

FIG. 2(b) displays simulation results of scattered light propagation through a "banana-shaped" region in tissue, simulated using MCmatlab [19] based on the MCXYZ model developed by Jacques and Li [20]. The simulation settings are detailed in Ref. [21]. Considering the SNR and flow sensitivity for speckle imaging in biomedical applications, typical exposure times range from 1 to 10 ms [37]. In our experiments, we used an exposure time of 1.64 ms (TINT_TBIN_ITERATIONS = 1024) with $\rho$ = 15 mm, resulting in an image acquisition rate of 361.7 fps.

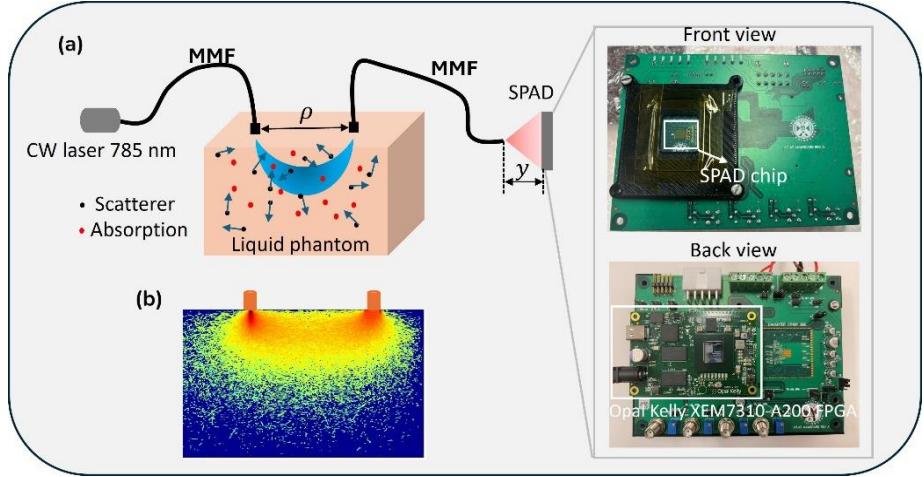

FIG. 2. (a) The schematic of the experimental setup illustrates the optical measurement system using multimode fibers (MMF) and a single-photon avalanche diode (SPAD). A continuous-wave (CW) laser (785 nm) propagates through a liquid phantom containing scatterers and absorbers, represented by black and red dots, respectively. y indicates the vertical position of the SPAD detector. Zoomed-in images show the SPAD chip (front view) and the Opal Kelly FPGA board (back view). (b) Simulation results depict scattered light traveling through a "banana-shaped" region in tissue. These simulations were conducted using MCmatlab [38] based on the model MCXYZ developed by Jacques and Li [39]. Detailed simulation settings can be found in Ref. [40].

**E. Data Processing**

The data processing pipeline is shown in FIG. 3. Raw speckle data, also known as intensity imaging data under PC mode, is initially acquired as a three-dimensional matrix. The first step involves identifying and removing bad pixels, which are detected based on irregular signal characteristics or abnormally high noise levels. Next, $\kappa$ is extracted to isolate the signal from noise. Shot noise reduction is then applied to minimize high-frequency fluctuations arising from the stochastic nature of photon detection, significantly enhancing the SNR. Finally, filtering is performed to smooth the signal and preserve essential features for further analysis. This processing pipeline ensures that the data is reliable and of high quality for subsequent interpretation and modeling.

The data processing pipeline for speckle contrast analysis follows a structured approach to ensure high-quality and reliable data extraction, as illustrated in FIG. 3. The process begins with the collection of raw speckle data, a three-dimensional matrix that records intensity fluctuations over time. Once the data is acquired, a pixel quality assessment is performed to identify and eliminate bad pixels that could introduce artifacts or bias into the analysis. Specifically, 'hot' pixels—those with significantly higher dark counts [23]—are detected and removed using a 3-sigma rule. Any pixel that deviates more than three standard deviations from the mean is replaced with NaN to prevent its influence on subsequent calculations, implemented with:

$$I[|I - \bar{I}| > 3\sigma] = NaN. \qquad (6)$$

where $I$ is the frame data, and $\bar{I}$ is the mean intensity of frame data, $\sigma$ is the standard deviation of the frame data.

Once bad pixels are identified, they are excluded from further calculations. For valid pixels, $\kappa$ is computed, and the BFi is derived using the inverse squared speckle contrast formula: BFi = $1/\kappa^2$. The resulting signal is then subjected to noise reduction algorithms to minimize fluctuations caused by external disturbances or system imperfections. Filtering techniques, including a smoothing window of 20, are applied to refine the data and highlight physiologically relevant blood flow variations. Finally, the processed data is visualized, enabling the interpretation of BFi dynamics over time. This pipeline ensures the extraction of accurate and meaningful blood flow information while minimizing the impact of noise and artifacts.

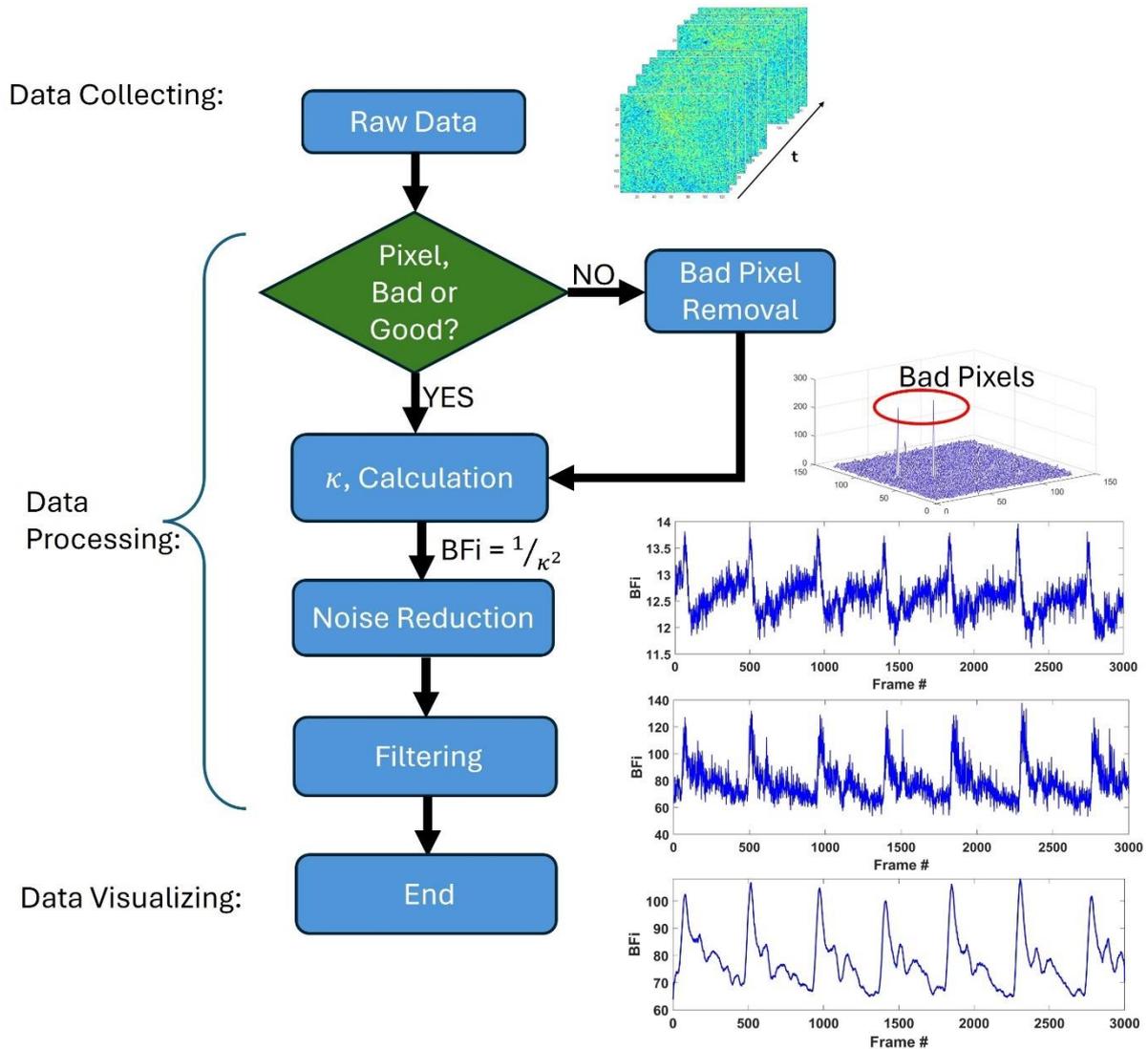

FIG. 3. The DSCA data analysis flow. Once the raw data is acquired, bad pixels (highlighted in red circles) are identified and removed, followed by $\kappa$ calculation, noise reduction, and final filtering. The final processed data is visualized for interpretation. The right-side images provide visual insights into raw data, bad pixel identification, and signal refinement at different stages.

## III. RESULTS

### A. Simulation Results

As a function of $\rho$ and $T$, the speckle contrast ($\kappa$) is computed using Eq. (2) and presented in Figs. 4(a) and 4(b). In FIG. 4(a), $\kappa$ decreases with increasing $\rho$ for $T$ = 1, 3, and 5 ms with a smaller $T$ exhibiting a higher $\kappa$. This trend aligns with the expectations that a larger $\rho$ leads to greater photon scattering and diffusion, thereby reducing $\kappa$. FIG. 4(b) demonstrates the nonlinear decrease of $\kappa$ and $T$ at various $\rho$ values. As $T$ increases, $\kappa$ decreases in a nonlinear fashion, with the rate of decrease being more pronounced at a smaller $\rho$, indicating that a shorter $T$ captures more high-frequency speckle fluctuations associated with BF. To further explore this dependency, FIG. 4(c) presents a 3D visualization of $\kappa$ as a function of $\rho$ and $T$. The surface plot clearly shows that $\kappa$ is highest at smaller $\rho$ and $T$, confirming the interplay between spatial and temporal factors in speckle contrast dynamics. Finally, FIG. 4(d) presents the relationship between $1/\kappa^2$ and $\alpha D_b$ (BFi), where it is evident that BFi increases linearly with $1/\kappa^2$. Unlike DCS, which requires fitting the second-order correlation function $g_2$, our direct relationship (BFi = $1/\kappa^2$) simplifies the calculation and improves accuracy and computational efficiency. The simulations were conducted when $\lambda$ = 785 nm, $\mu_a$ = 0.01 mm$^{-1}$, $\mu_s'$ = 1 mm$^{-1}$, and the refractive index $n$ = 1.33, representative of a typical tissue-mimicking medium.

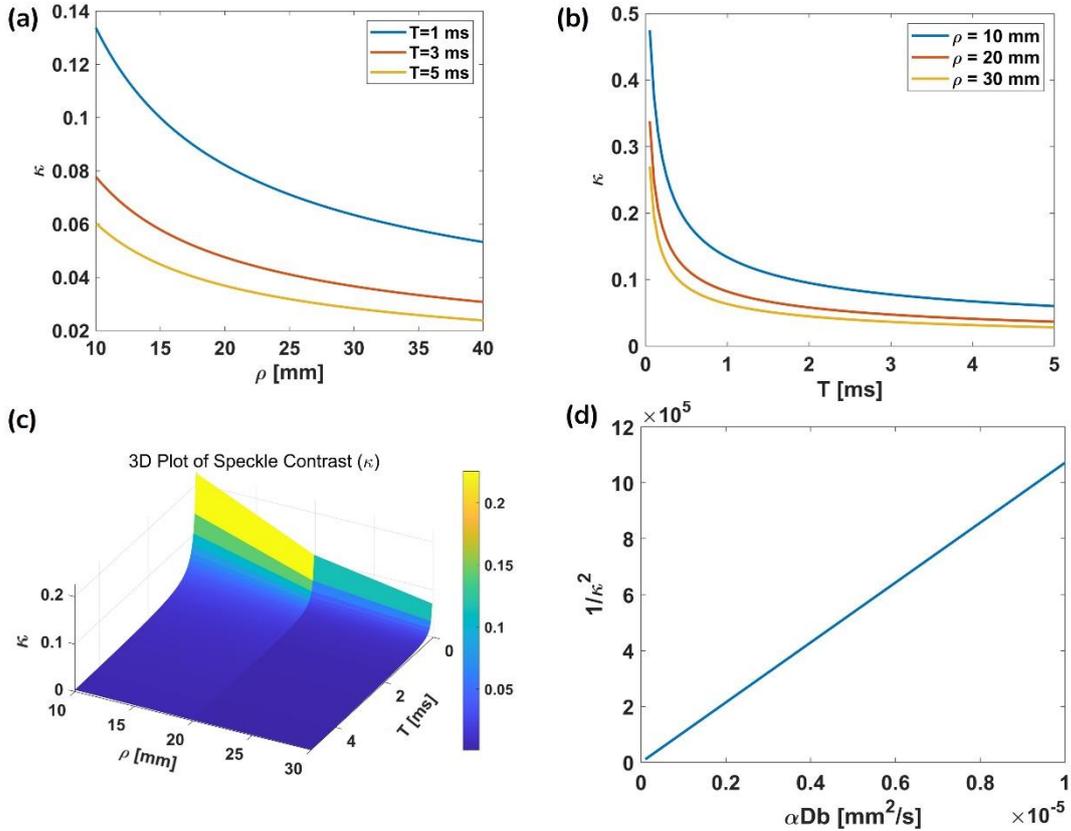

FIG. 4. Numerical simulations showing the relationship between $\kappa$ and various parameters. (a) $\kappa$ as a function of $\rho$ for $T = 1, 3, 5$ ms. (b) $\kappa$ as a function of $T$ for $\rho = 10, 20, 30\ mm$. (c) 3D visualization of $\kappa$ as a function of both ρ and $T$. (d) A linear relationship between of $1/\kappa^2$ and $\alpha D_b$, showing the expected theoretical trend, with $T$ = 2 ms. Here $\lambda = 785\ nm$, $\mu_a = 0.01\ mm^{-1}$, $\mu'_s = 1\ mm^{-1}$, and $n = 1.33$.

**B. Measurement Flexibility**

The measurement flexibility of the ATLAS-DSCA was assessed by varying the TINT_TBIN_ITERATIONS parameter in our custom software. Table II summarizes the relationship between TINT_TBIN_ITERATIONS, exposure time, frame readout time, frame time, and frame rate for the global shutter mode. As the exposure time increases, the frame rate decreases, highlighting the impact of a longer integration period on the data acquisition speed. Specifically, increasing TINT_TBIN_ITERATIONS leads to a proportional rise in the exposure time, significantly reducing the achievable frame rate. For instance, with a clock frequency of 20 MHz and TINT_TBIN_ITERATIONS set to 32, the system achieves a frame rate of 849.2 fps. However, at TINT_TBIN_ITERATIONS = 65,535, the frame rate drops to 9.4 fps.

Table II. Relationship between TINT_TBIN_ITERATIONS, exposure time, frame readout time, frame time, and frame rate under clock frequency = 20 MHz, TBIN_CLK_PERIODS = 32 and global shutter mode.

| TINT_TBIN_ITERATIONS | Exposure time (s) | Frame readout time (s) | Frame time (s) | Frame rate (fps) |
|---|---|---|---|---|
| 32 | 0.0000512 | 0.0011264 | 0.0011776 | 849.2 |
| 64 | 0.0001024 | 0.0011264 | 0.0012288 | 813.8 |
| 128 | 0.0002048 | 0.0011264 | 0.0013312 | 751.2 |
| 256 | 0.0004096 | 0.0011264 | 0.001536 | 651 |
| 512 | 0.0008192 | 0.0011264 | 0.0019456 | 514 |
| 1024 | 0.0016384 | 0.0011264 | 0.0027648 | 361.7 |
| 2048 | 0.0032768 | 0.0011264 | 0.0044032 | 227.1 |
| 4096 | 0.0065536 | 0.0011264 | 0.00768 | 130.2 |
| 8192 | 0.0131072 | 0.0011264 | 0.0142336 | 70.3 |
| 16384 | 0.0262144 | 0.0011264 | 0.0273408 | 36.6 |
| 32768 | 0.0524288 | 0.0011264 | 0.0535552 | 18.7 |
| 65535 | 0.1048560 | 0.0011264 | 0.1059824 | 9.4 |

FIG. 5 illustrates the impact of different TINT_TBIN_ITERATIONS values (32, 1024, and 4096) on normalized BFi measurements over the frame number. Each TINT_TBIN_ITERATIONS value corresponds to a different frame rate, as shown in Table II, which represents the sampling rate. For example, TINT_TBIN_ITERATIONS values of 32, 1024, and 4096 correspond to sampling rates of 849.2 Hz, 361.7 Hz, and 130.2 Hz, respectively. Fewer iterations result in a higher temporal resolution, allowing for the capture of rapid dynamics, whereas more iterations produce smoother curves due to the enhanced SNR (SNR = $10log_{10}(\frac{Signal\ power}{Noise\ power})$), albeit with a reduced temporal resolution. Here, we obtain SNR values of 12.10 dB, 25.13 dB, and 30.32 dB for TINT_TBIN_ITERATIONS = 32, 1024, and 4096, respectively. A lower TINT_TBIN_ITERATIONS means a shorter $T$, leading to a lower $\kappa$, which is in good agreement with the simulation results shown in FIG. 4(b). FIG. 5(b) presents the raw signal amplitude for the three iteration settings, highlighting the increased signal intensity and noise levels associated with lower iterations. In FIG. 5(c), the power spectral density (PSD) analysis, computed using Welch's method, further reveals the impact of iteration settings on signal frequency content. Welch's method can estimate the PSD by segmenting the signal into overlapping sections, applying a window function, and averaging the periodograms. The PSD estimate using Welch's method is given by:

$$PSD(f) = \frac{1}{M}\sum_{m=1}^{M} P_m(f), \quad P_m(f) = \frac{1}{N}|FFT(x_m)|^2, \tag{7}$$

where $f$ is the frequency at which the PSD is estimated, $M$ is the number of overlapping segments into which the signal is divided, and $P_m(f)$ is the periodogram (power spectrum for a given segment) of the $m$-th segment at $f$. $N$ is the length of each segment, and $x_m$ is the $m$-th segment of the signal. FFT stands for Fast Fourier Transform. By averaging multiple periodograms, Welch's method reduces variance and provides a smoother representation of the frequency components. The low iteration setting (red) allows for higher frequency components to be preserved, whereas increasing the iterations results in a more attenuated high-frequency response, favoring smoother signals. These findings emphasize the trade-off between the temporal resolution and signal quality, where fewer iterations improve temporal tracking but introduce higher noise, whereas more iterations enhance signal stability at the cost of high-frequency information. These findings highlight the trade-off between the temporal resolution and data quality, underscoring the system's versatility to meet various experimental needs. The measurements were conducted on the left arm of a healthy volunteer at $\rho$ = 12 mm (laser power: 4 mW), with the SPAD sensor set to a clock frequency of 20 MHz and TBIN_CLK_PERIODS = 32. The clock frequency is also adjustable (e.g., 50 MHz, 75 MHz) depending on experimental requirements.

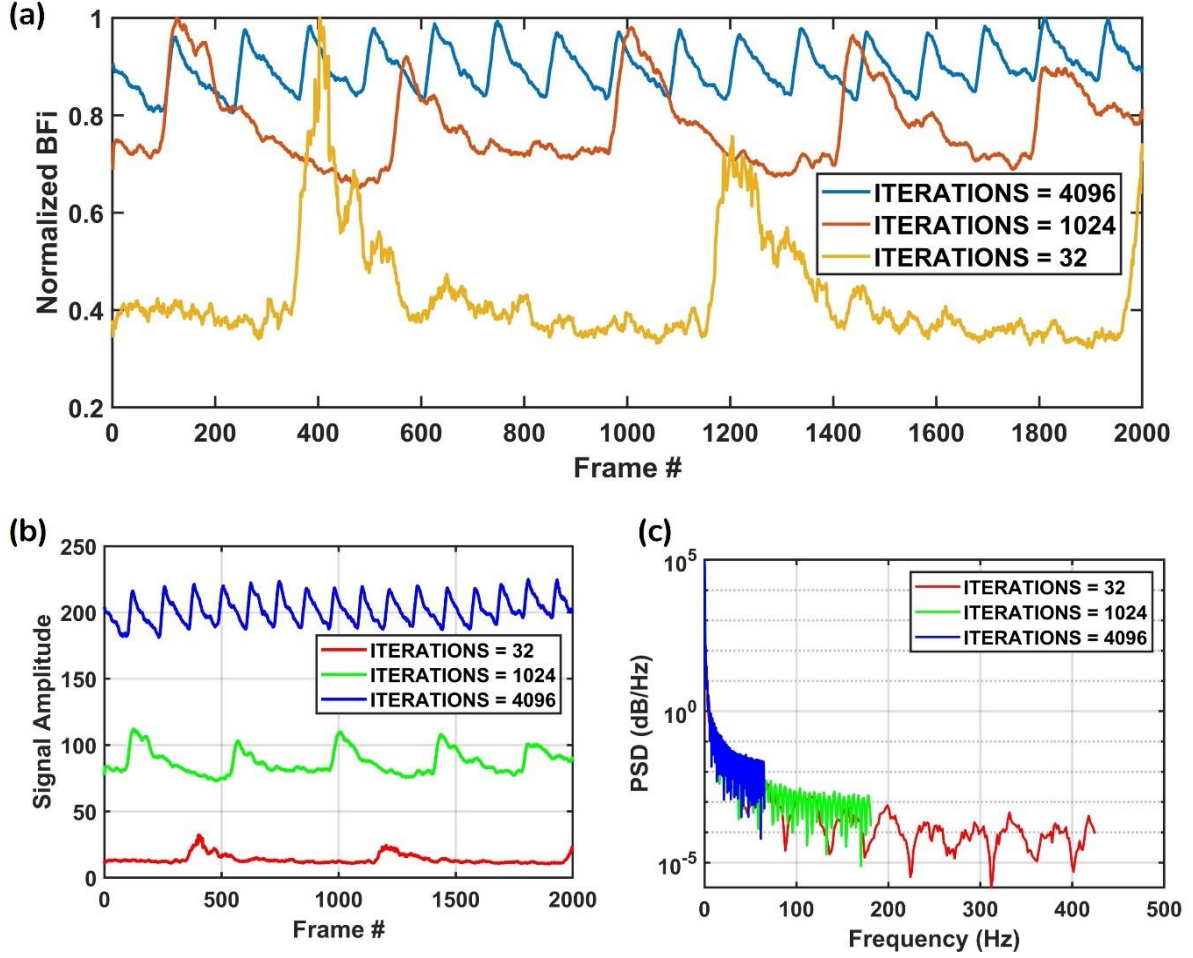

FIG. 5. (a) Normalized BFi measured over time with varying TINT_TBIN_ITERATIONS values (32, 1024, and 4096). (b) Raw signal amplitudes across different iteration settings, highlighting variations in signal intensity and noise levels. (c) The power spectral density (PSD) analysis of the signals, demonstrating the impact of iteration settings on frequency content, where lower iterations allow for higher frequency resolution whereas higher iterations result in smoother signals with reduced high-frequency components.

## C. Phantom and *In Vivo* Measurements

### 1. Phantom Measurements

Phantom measurements were performed using a setup that simultaneously compared the ATLAS-DSCA and a conventional CMOS-DSCA systems. We employed a home-made solid silica phantom (with unmeasured optical properties, as no quantitative calculation was intended) containing a vibration motor whose intensity can be controlled externally. FIG. 6(a) shows the experimental setup, where a laser illuminates the solid silica phantom through an MMF, and scattered light is collected via two separate MMFs: one connected to the ATLAS-DSCA system and the other to the CMOS-DSCA system. A vibration motor embedded in the phantom induces controlled motion, with vibration levels (0, 12, 24, 36, 52) adjusted through an external controller. FIG. 6(b) presents the time-dependent normalized BFi measured by both systems. The stepwise increases in BFi correspond to different vibration levels applied to the phantom, illustrating the responsiveness of ATLAS-DSCA and CMOS-DSCA to dynamic changes. A noticeable discrepancy between the BFi values obtained from the two systems likely reflects inherent differences in their measurement characteristics, such as quantum efficiency, dark current, and other detector-specific factors, between CMOS and SPAD detectors. This discrepancy validates the feasibility of the ATLAS-DSCA method under controlled conditions, paving the way for its application in complex in vivo scenarios.

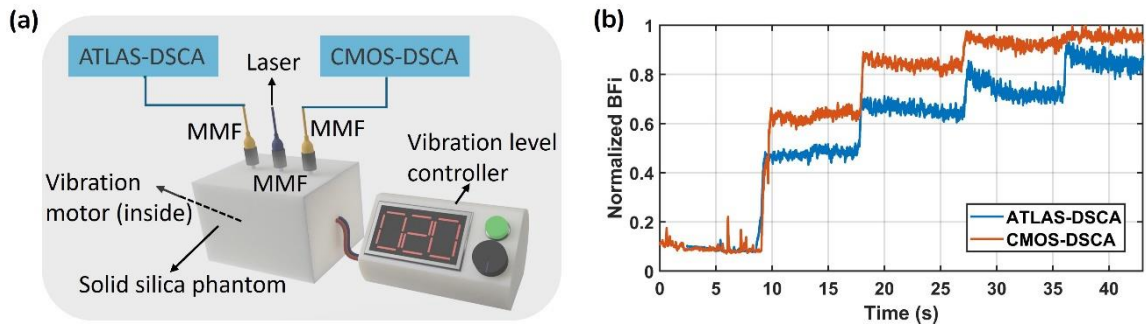

FIG. 6. (a) Experimental setup for DSCA: A laser illuminates a solid silica phantom via an MMF, with scattered light collected by two separate MMFs for ATLAS- and CMOS-DSCA detection. A vibration motor inside the phantom generates controlled motion at intensities 0,12,24,35,52. (b) Normalized BFi over time for ATLAS-DSCA (blue) and CMOS-DSCA (orange), showing responses to varying vibration levels.

## 2. Arm Cuff Occlusion

To further demonstrate the ATLAS-DSCA system, *in vivo* measurements were conducted using an arm cuff occlusion model at $\rho = 10, 20,$ and $30$ mm. As shown in FIG. 7(a), volunteers sat comfortably with their left arm placed on a pad and the ATLAS-DSCA probe attached to the wrist. FIG. 7(b) presents relative BFi (rBFi = BFi / $BFi_{baseline}$) time series data for four subjects at different $\rho$ values, highlighting the microvascular hemodynamic changes during occlusion and subsequent recovery. After a baseline period of approximately 3.5 s, a blood pressure cuff was inflated to 200 mmHg for 7 s. Upon cuff release, blood flow recovery was recorded for 4 s. The gray-shaded regions indicate the cuff inflation period, during which a marked decrease in rBFi is observed. Upon cuff release, the rBFi overshoots to a hyperemic value much larger than the baseline as expected, with varying recovery kinetics across subjects and measurement depths. Notably, the reduction in rBFi is more pronounced at a shorter $\rho$ ($\rho = 10$ mm), suggesting that superficial microvascular networks exhibit a stronger response than deeper tissues, where SNR may be lower.

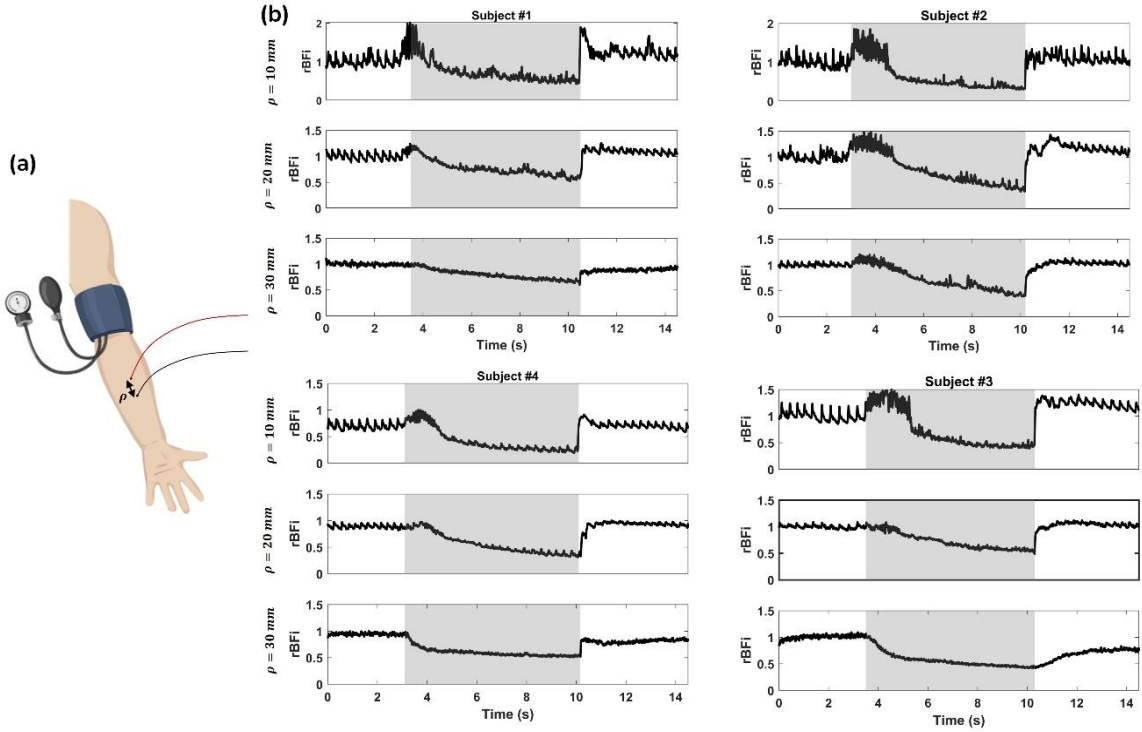

FIG. 7. (a) The schematic representation of the experimental setup, where a blood pressure cuff is placed on the upper arm to induce controlled vascular occlusion, with a certain $\rho$. (b) Temporal rBFi variations at $\rho$ = 10, 20, and 30 mm for four subjects (Subjects #1~#4). The gray-shaded regions indicate the period of cuff inflation, during which blood flow is restricted. The signals show a characteristic decrease in rBFi during occlusion, followed by a recovery phase after cuff deflation.

### 3. *In Vivo* Forehead Measurements

ATLAS-DSCA was also applied to monitor cerebral blood flow on the adult human forehead, as shown in FIG. 8(a). FIG. 8(b) represents the temporal variations in normalized BFi (blue) and photoplethysmography (PPG, red) signals at $\rho$ = 20 mm and $\rho$ = 25 mm. The BFi reflects cerebral blood flow fluctuations, whereas the PPG indicates peripheral blood volume changes. A clear phase shift between the two signals is observed, highlighting the differences in hemodynamic responses at varying tissue depths. The average intensity, which is similar to the PPG signal, serves as an indicator of blood volume [41]. In our measurements, the BF waveform exhibits sharper peaks and more detailed features within each cardiac cycle than the PPG waveform, with the BF peak consistently preceding the blood volume peak – a phenomenon also reported in Ref. [16].

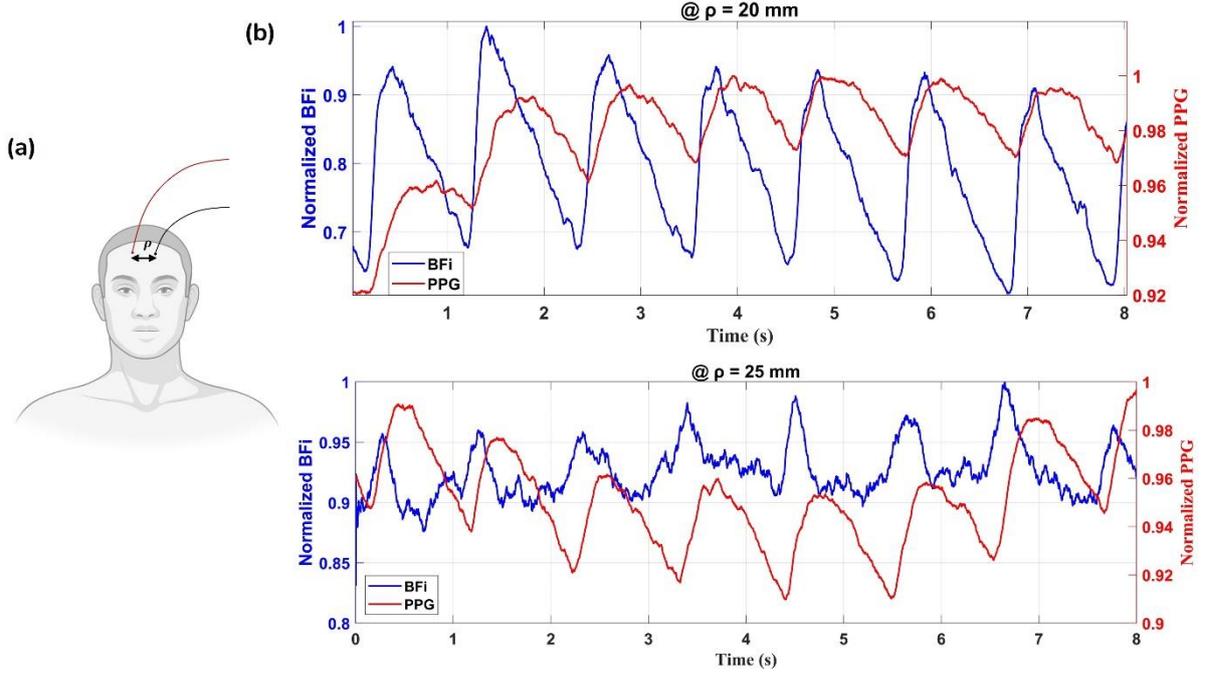

FIG. 8. (a) The schematic representation of the measurement setup. (b) Normalized BFi (in blue) and photoplethysmography (PPG, in red) signals at $\rho$ = 20 and 25 mm. The signals exhibit a clear correlation, reflecting hemodynamic fluctuations over time.

To further demonstrate the sensitivity of ATLAS-DSCA, FIG. 9 presents cerebral function monitoring during a mental arithmetic task (at $\rho$ = 20 mm) in two subjects. FIG. 9(a) shows the experimental setup, where the probe was placed on the forehead, targeting the prefrontal cortex, which plays a key role in cognitive processes such as reading unfamiliar text, planning, and working memory [42,43]. In this experiment, subjects rested for 8 seconds before being presented with math questions for 30 seconds, followed by a recovery period after removing the questions. FIG. 9(b) illustrates a representative BFi measurement from one subject. The upper and lower envelopes of the BFi signal (denoted as up1 and lo1, respectively) were calculated using MATLAB's envelope function with the 'peak' method (window size = 250). The trend of the BFi signal is defined as:

$$BFi_{signal} = \frac{up1+lo1}{2}. \qquad (8)$$

Taking the first 8 seconds as the baseline, the percentage change in BFi is calculated as:

$$\text{BFi changes} = \frac{|BFi_{signal}-BFi_{baseline}|}{BFi_{baseline}} \times 100\%. \qquad (9)$$

As shown in FIG. 9(c) and 9(d), we observed that blood flow significantly increased by 3.8% to 10.1% (average over three trials per subject) during activation and returned to baseline values post-activation, consistent with expected physiological responses. The shaded area represents the standard deviation of the three trials for each subject, calculated using built-in MATLAB functions. These findings demonstrate the feasibility of using ATLAS-DSCA to track neurovascular dynamics in response to cognitive stimuli, offering valuable insights into cerebral hemodynamic responses associated with mental effort.

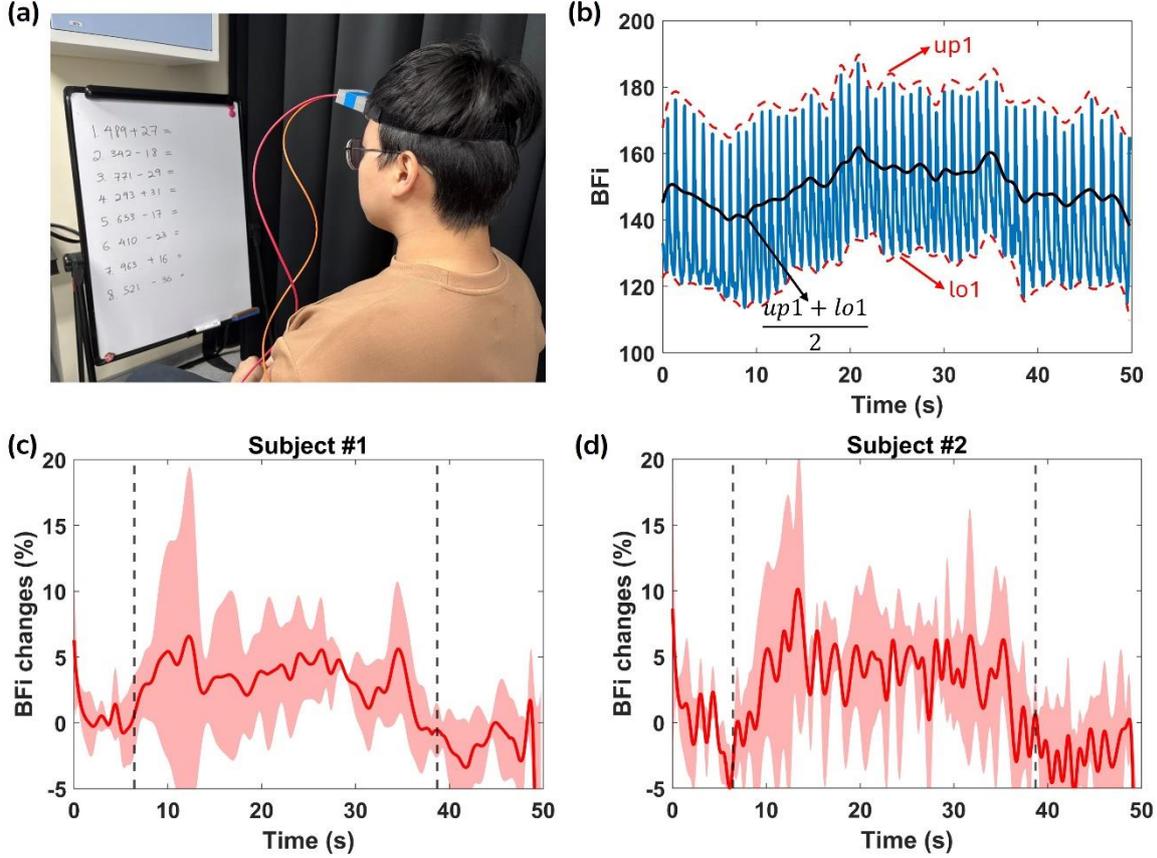

FIG. 9. (a) The experimental setup showing a subject performing mental arithmetic tasks while cerebral blood flow was monitored using a home-made head-mounted optical probe. (b) The recorded BFi (in blue) during the task. The red dotted lines represent the upper and lower envelopes (up1 and lo1, respectively), computed using the MATLAB envelop function with the 'peak' method. The black line represents the average of up1 and lo1, providing a smoothed trend of BFi signal. (c) & (d) Relative changes in BFi (%) for Subject #1 and Subject #2, respectively. The shaded region represents the standard deviation, whereas the dashed vertical lines indicate the start and end of the cognitive task.

## IV. DISCUSSION

We introduce the ATLAS-DSCA system, which leverages a 512×512 SPAD array for non-invasive deep tissue blood flow measurement. Our findings show that the ATLAS-DSCA offers a comparable performance to traditional CMOS-DSCA systems, but with the added benefit of a higher sampling rate. Validation studies, including arm cuff occlusion, *in vivo* forehead measurements, and cognitive task trials in healthy subjects, all confirm the robustness and effectiveness of our approach.

Phantom experiments (FIG. 6(b)) conducted alongside a CMOS-DSCA system further validate the reliability of the ATLAS-DSCA. Additionally, arm cuff occlusion studies (FIG. 7(b)) highlight the system's ability to assess deep tissue blood flow, demonstrated by the pronounced hyperaemic response following cuff release. Cognitive task experiments (FIG. 9(c) and 9(d)) also show a marked increase in blood flow during mental arithmetic, aligning with previous research [28,44].

As presented in Table I, Kim *et al.*'s SCOS system achieved $\rho = 40$ mm with a sampling rate of 160 Hz [28]. However, their approach involves certain trade-offs, such as the use of lower-cost CMOS cameras, which introduce higher readout noise, reduced bit depth, and non-linear or non-uniform camera gains. To improve photon flux without exceeding safety limits, they implemented a pulsing strategy (10% duty cycle) with an optical chopper on a volume holographic laser, resulting in a more complex optical setup. Additionally, their system incorporates a complicated 4f structure, making it bulkier and less portable.

In contrast, Huang *et al.*'s system [29] achieved $\rho = 50$ mm with a sampling rate of 80 Hz, which captures low-frequency signals (< 65 Hz), unable to resolve high-frequency components, as shown in FIG. 5(c) (blue curve), which may result in missing critical information. However, our ATLAS-DSCA system is capable of resolving high-frequency signals (> 200 Hz), making it ideal for analyzing dynamic, rapidly changing physiological signals. They used a higher laser power (45 mW) to achieve $\rho = 50$ mm. Their detection probe is designed to place CMOS image sensors close to the patient's skin, eliminating the need for detection fibers. This increases photon collection and improves the SNR, making it ideal for adult stroke diagnostics. However, this design is less suitable for premature or neonatal patients, who require minimal disruption.

In comparison, our ATLAS-DSCA system, while achieving $\rho = 30$ mm, distinguishes itself with an exceptional sampling rate exceeding 800 Hz. Not only can it detect human blood flow, but it is also suitable for monitoring animal health and wellbeing (for small animals, the heart rate typically ranges from 4 to 12.5 Hz, or 240 to 750 beats per minute [45]). This high sampling rate also opens the door to broader applications, including ultrasound detection, assuming the SPAD's bandwidth and sensitivity support higher frequencies. A major advantage of our system is its flexibility: the ATLAS can incorporate multiple fibers by dividing the sensor into several sensing clusters, offering greater versatility. Additionally, the ATLAS-DSCA system can operate in DCS mode, providing $\rho = 50$ mm (research results to be published separately) using on-chip autocorrelators [24,46].

Several studies [26,47–49] have compared DCSA/SCOS and DCS technologies, employing either Monte Carlo simulations or separate detectors—one dedicated to DCSA/SCOS and another to DCS. In contrast, our DSCA/DCS system allows switching between DCS and DSCA modes, facilitating a more controlled and direct comparison.

Furthermore, there are two main approaches for capturing temporal dynamics: temporal sampling methods (such as DCS using SPAD arrays) and speckle ensemble methods (like SVS/DSCA and LSCI). While temporal sampling methods require extremely high frame rates (above 26 kfps, as highlighted in Wang *et al.*'s review [34]) to capture fast fluctuations, speckle ensemble methods improve SNR by integrating over multiple speckles. Our work demonstrates that, although SPAD sensors may not be ideal for traditional DCS (calculating $g_2$ using sequential frame data), they hold significant promise in DSCA, SCOS, and SVS applications.

Lastly, it is important to note that our SPAD array maintains a high SNR even with very short exposure times (as low as 51.2 µs), thanks to its lack of readout noise and minimal dead time—critical advantages that enhance the performance of the ATLAS-DSCA system.

We believe, recent advancements in SPAD technology—such as the development of large-area SPAD arrays (512×512) and improved time-gated detection—suggest that SPADs could play a more prominent role in next-generation DSCA/SCOS applications. Their unparalleled temporal resolution, high sampling rate and ability to perform time-gated measurements make them particularly attractive for deep-tissue blood flow imaging and functional neuroimaging applications. Moreover, the algorithm we developed for the fiber-based ATLAS-DSCA system (see FIG. 3) may server as an effective foundation for advancing SPAD-based DSCA/SCOS methodologies.

## V. CONCLUSION

In conclusion, we present a groundbreaking fiber-based, ultra-high-speed DSCA system using a large-format SPAD camera for non-invasive deep tissue blood flow sensing. By combining a custom-designed SPAD array with cutting-edge optical and signal processing techniques, our system achieves an exceptional temporal resolution and sensitivity, enabling the capture of rapid, dynamic blood flow variations across a broad field of view. Extensive experimental validations—from phantom studies and cuff occlusion tests to in vivo measurements—demonstrate that our DSCA system reliably detects deep tissue hemodynamic changes and outperforms traditional CMOS-based systems in sampling rate.

Moreover, the system's high-speed capabilities open the door to a broad range of biomedical applications, from human clinical diagnostics to animal physiological monitoring, with potential for expanding into ultrasound detection. While this work underscores the transformative potential of large SPAD array cameras in DSCA, we anticipate that future sensor technology and signal processing advancements will further elevate its performance and broaden its clinical and research utility.

## ACKNOWLEDGMENTS

This work has been funded by the Engineering and Physical Sciences Research Council (Grant No. EP/T00097X/1 and No. EP/T020997/1): the Quantum Technology Hub in Quantum Imaging (QuantiC) and the University of Strathclyde, as well as A*STAR Biomedical Engineering Programme (BEP) C221318003 and A*STAR BMRC CRF fund 2024.

## DATA AVAILABILITY

The data that support the findings of this study are available from the corresponding author upon reasonable request.

## DECLARATIONS

**Ethics Approval Statement** Six healthy participants, aged between 20 and 35 years, with no history of neurological disorders, were recruited for this study. Demographic factors such as sex, gender, race, and ethnicity were not criteria for recruitment. Participants were selected through internal department advertisements. Of the six, four were involved in cuff occlusion measurements, while two were selected to measure changes in cerebral blood flow (CBF) induced by behavioral variations in the prefrontal cortex during mental arithmetic tasks. All experimental procedures and protocols were approved by the Institutional Review Board at

A*STAR and adhered to their guidelines. Prior to participation, each subject provided written informed consent.

**Conflict of Interests** The authors declare no competing interests.

## APPENDIX: DERIVATION OF SPECKLE CONTRAST EXPRESSION USING BASED ON FIELD AUTOCORRELATION.

Substituting $g_1(\rho, \tau) = G_1(\rho, \tau)/G_1(\rho, 0)$ into Eq. (2), the speckle contrast expression can be reformulated as

$$\kappa^2(\rho, T) = \frac{8\beta}{(FT)^2 G_{.0}^2} \sum_{i=1}^{2} \sum_{j=1}^{2} \frac{(-1)^{i+j}}{r_i r_j (r_i + r_j)^4} \times [X_{ij}(T) - X_{ij}(0) + Y_{ij}FT], \qquad (10)$$

where $X_{ij}(T) = [(r_i + r_j)^2 (K_0^2 + FT) + 3(r_i + r_j)\sqrt{K_0^2 + FT} + 3] \times e^{-(r_i+r_j)\sqrt{K_0^2+FT}}$,

$Y_{ij} = \frac{1}{2}(r_i + r_j)^2[1 + K_0(r_i + r_j)] \times e^{-K_0(r_i+r_j)}$. For full details of the derivation, please refer to Ref. [50]. Eq. (10) is a general formular describing the behavior of $\kappa$ with respect to the exposure time $T$ and $\rho$. In our theoretical simulation, this equation is used to calculate the corresponding speckle contrast.